\documentclass{PoS}

\usepackage{graphicx}
\usepackage{amsmath}
\usepackage{amssymb}

\newcommand{\be}{\begin{equation}}
\newcommand{\ee}{\end{equation}}
\newcommand{\bea}{\begin{eqnarray}}
\newcommand{\eea}{\end{eqnarray}}
\newcommand{\bi}{\begin{itemize}}
\newcommand{\ei}{\end{itemize}}
\newcommand{\ben}{\begin{enumerate}}
\newcommand{\een}{\end{enumerate}}
\newcommand{\bt}{\begin{tabbing}}
\newcommand{\et}{\end{tabbing}}
\newcommand{\non}{\nonumber}

\title{
Tensor network approach to real-time path integral
}

\ShortTitle{
Tensor network approach to real-time path integral
}

\author{
   \speaker{Shinji Takeda}
   \\
Institute of Physics, Kanazawa University, Kanazawa 920-1192, Japan\\
         E-mail: \email{takeda@hep.s.kanazawa-u.ac.jp}}
   
\abstract{
We present a tensor network representation of the path integral for the one-component real scalar field theory
in 1+1 dimensional Minkowski space-time.
It is numerically verified by comparing with the
exact result in the non-interacting case.
}

\FullConference{37th International Symposium on Lattice Field Theory - Lattice2019\\
		16-22 June 2019\\
		Wuhan, China}

\begin{document}

\section{Introduction}

The real-time dynamics is one of the most important topics in modern physics,
however, the field theoretical approaches based on the path integral representation
suffer from the severe sign problem.
So far several approaches have been proposed to overcome the sign problem.
For example, the complex Langevin method is used to study
$\phi^4$ scalar field theory \cite{Berges:2005yt} and
the SU(2) gauge theory \cite{Berges:2006xc} in 3+1 dimensional Minkowski space-time.
In the course of studies, however, it turns out that the convergence issue
gets harder when the time extent is taken relatively larger.
Recently, the generalized Lefschetz thimble algorithm is
used in the study of the 1+1 dimensional $\phi^4$ theory \cite{Alexandru:2017lqr}
with the Schwinger-Keldysh setup \cite{Schwinger:1960qe,Keldysh:1964ud}.
This approach, however, also has a limitation on the lattice size and it seems hard to perform larger volume simulations.

Tensor network method can also avoid the sign problem and is another promising approach.
The idea of tensor networks was invented in the condensate matter physics and developed
with a help of quantum information \cite{White:1992zz,Ostlund:1995zz,Verstraete:2004cf,Schollwock:2005zz}.
Recently, the idea and technique have been used in elementary particle physics.
It is known that within a framework of the tensor networks there are two types of approaches:
the Hamiltonian approach and the Lagrangian one.
The former one
uses a tensor network in an ansatz form of a trial wave function for the variational method,
and the real-time dynamics is studied in various two dimensional models \cite{Pichler:2015yqa,Kuhn:2015zqa,Lin}. 
The letter approach is nothing but the path integral and we will use it in the following.
The aim of the study here is to demonstrate how to use the tensor network technique when evaluating the real-time path integral.
As a first trial, we work on
1+1 dimensional real scalar field theory with the Minkowskian metric.
In the following, we assume the lattice units $a=1$.

\section{Tensor network representation}
We consider 1+1 dimensional square lattice system 
whose coordinate is denoted by $x=(x_0,x_1)$ with $x_0=0,1,2,...,T-1$ and $x_1=0,1,2,...,L-1$ for time and space respectively.
In the following $T=L$ is assumed.
The boundary condition for both directions is taken to be periodic just for a simplicity,
though this is not a physically meaningful setting.
The one-component real scalar fields on the lattice are denoted by $\phi_x\in\mathbb{R}$.

The lattice action for the 1+1 dimensional real scalar $\phi^4$ theory with the Minkowskian metric is given by
\bea
S[\phi]
&=&
\sum_x
\left[
\frac{1}{2}
(\phi_{x+\hat0}-\phi_x)^2
-
\frac{1}{2}
(\phi_{x+\hat1}-\phi_x)^2
-
V(\phi_x)
\right]
\eea
with the potential term
\be
V(\phi)
=
\frac{1}{2}
(m_0^2-i\varepsilon) \phi_x^2
+
\frac{\lambda}{4!}\phi_x^4,
\ee
where $m_0, \lambda\in\mathbb{R}$ are the bare parameters  and
the Feynman prescription parameter $\varepsilon$ ($\varepsilon\in\mathbb{R}$, $\varepsilon>0$) is introduced and this plays
an important role when forming the tensor network representation as we will see later.
In an actual computation, note that we
will have to take $\varepsilon\to0$ limit for a bare physical quantity. 

In order to use the tensor network method,
first of all, one has to rewrite the path integral in terms of the tensor network representation.
For that purpose, let us rewrite it as a product of local factors,
\bea
{\cal Z}
&\equiv&
\int [d\phi]e^{iS[\phi]}
=
\int [d\phi]
\prod_{x}
H^{(0)}(\phi_x,\phi_{x+\hat0})
H^{(1)}(\phi_x,\phi_{x+\hat1}),
\label{eqn:hoppingfactor}
\eea
where the local factors
for the time-direction $H^{(0)}$ and for the space $H^{(1)}$ are explicitly given by\footnote{
Note that the phase factors $\exp(i\phi_x^2/2)$ associated with the kinetic terms has canceled in eq.(\ref{eqn:H0first}) and (\ref{eqn:H1first}).
}
\bea
H^{(0)}(\phi,\phi^\prime)
&=&
\exp
\left[
-i\phi\phi^\prime
-
\frac{i}{4}
V(\phi)
-
\frac{i}{4}
V(\phi^\prime)
\right],
\label{eqn:H0first}
\\
H^{(1)}(\phi,\phi^\prime)
&=&
\exp
\left[
+i\phi\phi^\prime
-
\frac{i}{4}
V(\phi)
-
\frac{i}{4}
V(\phi^\prime)
\right].
\label{eqn:H1first}
\eea
They are considered as a two-variable function.
An important property of the local factors is that
thanks to the presence of the Feynman prescription parameter
they are Hilbert-Schmidt operator
\be
\int_{-\infty}^{\infty} d\phi d\phi^\prime |H^{(\mu)}(\phi,\phi^\prime)|^2
<
\infty
\hspace{10mm}
\mbox{for}
\hspace{5mm}
\mu=0,1.
\ee
This fact immediately means that they are compact operator
and can be expanded as\footnote{
One can show that the singular value is independent of $\mu=0,1$.
Moreover, it can be shown that it is independent of the bare parameters $m_0$ and $\lambda$ since
they are included in the phase factor in eq.(\ref{eqn:H0first}) and (\ref{eqn:H1first}).
Therefore the singular value is a function of only $\varepsilon$, while
the bare parameter dependence are captured in the basis functions.
}
\be
H^{(\mu)}(\phi,\phi^\prime)
=
\sum_{k=0}^\infty
\Psi_k^{(\mu)}(\phi)\,\sigma_k\, \Phi_k^{(\mu)\ast}(\phi^\prime),
\label{eqn:general}
\ee
where $\Psi$ and $\Phi$ are orthonormal basis function satisfying
\bea
\int_{-\infty}^{\infty} d\phi\, \Psi_m^\ast(\phi)\, \Psi_n(\phi)
&=&
\int_{-\infty}^{\infty} d\phi\, \Phi_m^\ast(\phi)\, \Phi_n(\phi)=\delta_{mn},
\label{eqn:orthonormal}
\\
\sum_{k=0}^\infty \Psi_k(\phi)\,\Psi_k^\ast(\phi^\prime)
&=&
\sum_{k=0}^\infty \Phi_k(\phi)\,\Phi_k^\ast(\phi^\prime)
=
\delta(\phi-\phi^\prime),
\label{eqn:complete}
\eea
and $\sigma_k$ represents singular value and is non-negative \cite{Lay,Shimizu:2012zza}.

As a next step let us see how to obtain the basis functions and the singular values.
First, we expand $H^{(\mu)}$
\be
H^{(\mu)}(\phi,\phi^\prime)
=
\sqrt{2\pi}
\sum_{m,n=0}^\infty
\psi_m(\phi)\,X^{(\mu)}_{mn}\,\psi_n(\phi^\prime)
\label{eqn:Hmu}
\ee
in terms of the Hermite function $\psi_m$ which is the eigen-function
of the harmonic oscillator
\be
\psi_n(x)
=
\frac{1}{\sqrt{\pi^{1/2}n!2^n}}
H_n(x)e^{-x^2/2},
\hspace{10mm}
H_n(x):\mbox{Hermite polynomials},
\ee
and satisfies the orthonormal and the completeness relations as in eq.(\ref{eqn:orthonormal}) and (\ref{eqn:complete}).
Basically $H^{(\mu)}$ are an oscillating function thus it is reasonable to use $\psi_m$ as a basis
since the latter is also such a function.
By using the orthonormal property of $\psi_m$, the coefficient matrix $X_{mn}^{(\mu)}$, which is a complex valued symmetric matrix, is obtained as
\be
X^{(\mu)}_{mn}
=
\frac{1}{\sqrt{2\pi}}
\int_{-\infty}^\infty d\phi d\phi^\prime
\psi_m(\phi)H^{(\mu)}(\phi,\phi^\prime)\psi_n(\phi^\prime).
\ee
An evaluation of the double integral for such an oscillating function is rather technically demanding,
thus here we use a trick in the following.
In order to separate the integration into two parts, we use a formula\footnote{
This may be derived from $e^{ixp}=\sqrt{2\pi}\langle x|p\rangle=\sqrt{2\pi}\sum_{n=0}^\infty \langle x|n\rangle\langle n|p\rangle$
and the Fourier transformation of the Hermite function.}
\be
e^{\mp i\phi\phi^\prime}
=
\sqrt{2\pi}
\sum_{n=0}^\infty
(\mp i)^n
\psi_n(\phi)\psi_n(\phi^\prime).
\label{eqn:epp}
\ee
By using the formula together with eq.(\ref{eqn:H0first}) and (\ref{eqn:H1first}), $X^{(\mu)}_{mn}$ is given by
\be
X^{(\mu)}_{mn}
=
\sum_{k=0}^\infty
(-1)^{k\delta_{\mu0}} i^k
G_{mk}
G_{nk}
\hspace{5mm}
\mbox{with}
\hspace{5mm}
G_{mn}
=
\int_{-\infty}^\infty d\phi
\psi_m(\phi)
\psi_n(\phi)
\exp\left[-\frac{i}{4}V(\phi)\right].
\ee
The resulting single integration for $G_{mn}$, which is again
a complex valued symmetric matrix, can be exactly carried out for the free case
and then $G_{mn}$ is given by the hypergeometric function, while
for the interacting case it may be estimated by some numerical integration scheme
with a suitable deformation of integration path.
After obtaining $X_{mn}^{(\mu)}$ numerically, we apply SVD to it,
\be
X^{(\mu)}_{mn}
=
\sum_{k=0}^\infty
U^{(\mu)}_{mk}\sigma_k(V^{(\mu)\dag})_{kn},
\hspace{5mm}
\mbox{with}
\hspace{5mm}
U^{(\mu)}, V^{(\mu)}
\mbox{ : unitary matrix}.
\ee
By inserting the above equation into eq.(\ref{eqn:Hmu}), $H^{(\mu)}$ are given as
\bea
H^{(\mu)}(\phi,\phi^\prime)
&=&
\sqrt{2\pi}
\sum_{m,n=0}^\infty
\psi_m(\phi)
\left(
\sum_{k=0}^\infty
U^{(\mu)}_{mk}\sigma_k (V^{(\mu)\dag})_{kn}
\right)
\psi_n(\phi^\prime)
\non\\
&=&
\sqrt{2\pi}
\sum_{k=0}^\infty
\left(
\sum_{m=0}^\infty
\psi_m(\phi)
U^{(\mu)}_{mk}
\right)
\sigma_k
\left(
\sum_{n=0}^\infty
(V^{(\mu)\dag})_{kn}
\psi_n(\phi^\prime)
\right)
\non\\
&=&
\sqrt{2\pi}
\sum_{k=0}^\infty
\Psi^{(\mu)}_k(\phi)
\sigma_k
\Phi^{(\mu)\ast}_k(\phi^\prime),
\label{eqn:Hpsp}
\eea
where in the last step, we have defined new basis functions
\bea
\Psi^{(\mu)}_k(\phi)
&=&
\sum_{m=0}^\infty
\psi_m(\phi)
U^{(\mu)}_{mk},
\hspace{10mm}
\Phi_k^{(\mu)\ast}(\phi)
=
\sum_{n=0}^\infty
(V^{(\mu)\dag})_{kn}
\psi_n(\phi).
\eea
In this way, the basis functions and the singular values are obtained.
In an actual calculation, one cannot deal with the infinite range of index for the matrix,
thus one has to truncate it and should monitor the truncation errors of some physical quantities.
Note that the singular values of $H^{(\mu)}$ are the same as those of $X^{(\mu)}_{mn}$.

Final step is to obtain the tensor.
By collecting all building blocks, the tensor is formed as
\bea
T_{ijkl}
&=&
2\pi\sqrt{\sigma_i\sigma_j\sigma_k\sigma_l}
\int_{-\infty}^\infty d\phi
\Psi_i^{(0)}(\phi)
\Psi_j^{(1)}(\phi)
\Phi_k^{(0)\ast}(\phi)
\Phi_l^{(1)\ast}(\phi).
\label{eqn:tensor}
\eea
Note that the numbering of the basis function turns out to be an index of tensor $i,j,k$ and $l$.
Basically, the range of the indices is infinity 
but to put it on a computer one has to truncate it.
The truncation order is denoted as $N$
and the index range is restricted as $0\le i,j,k,l\le N$ in the following.
The integration in eq.(\ref{eqn:tensor}) can be carried out exactly and it is given by
the hypergeometric function again.


\section{Numerical results}

\begin{figure}[t]
\begin{center}
\begin{tabular}{cc}
\scalebox{0.9}{\includegraphics{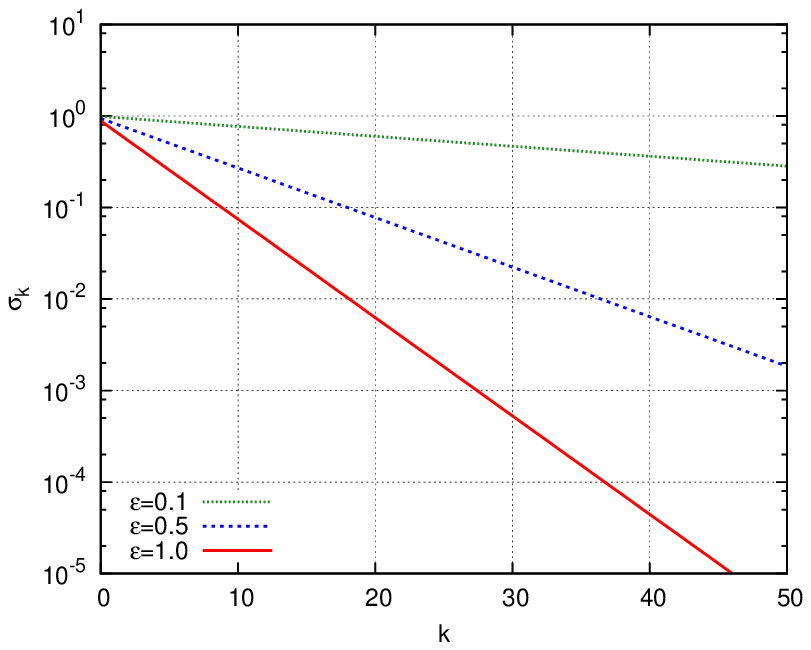}}
&
\hspace{-15mm}
\scalebox{0.9}{\includegraphics{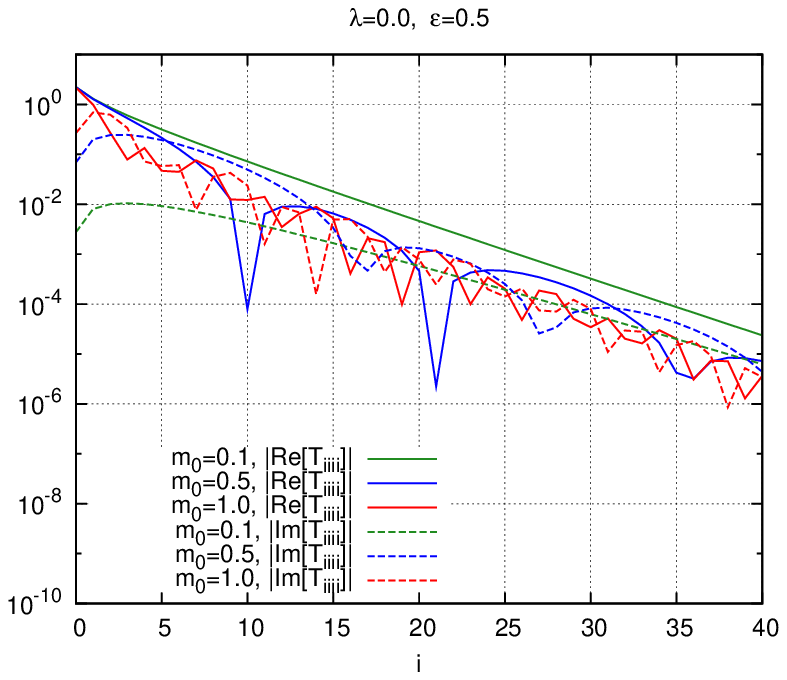}}
\end{tabular}
\end{center}
\vspace{10mm}
\caption{
Left panel: Singular values $\sigma_k$ of $H^{(\mu)}(\phi,\phi^\prime)$ (or equivalently $X^{(\mu)}_{mn}$) with
$\varepsilon=0.1$, $0.5$ and $1$.
Right panel: The absolute value of ${\rm Re}[T_{iiii}]$ and the imaginary part at
$\lambda=0$ and $\varepsilon=0.5$ with various $m_0$ values.
}
\label{fig:SV}
\end{figure}

The resulting singular values $\sigma_k$ are shown in Fig.~\ref{fig:SV} (left).
For large $\epsilon$ where a damping in the integrand gets stronger, the hierarchy
is sharper as expected, while
for small $\epsilon$, the hierarchy is not clear and
the information compression is not well realized.

To see the hierarchy structure of the tensor elements, $T_{iiii}$ are shown in Fig.~\ref{fig:SV} (right).
Basically the size of elements tends to be exponentially small for larger $i$.
This behavior allows us to truncate the tensor index.

\begin{figure}[t]
\begin{center}
\begin{tabular}{cc}
\scalebox{0.9}{\includegraphics{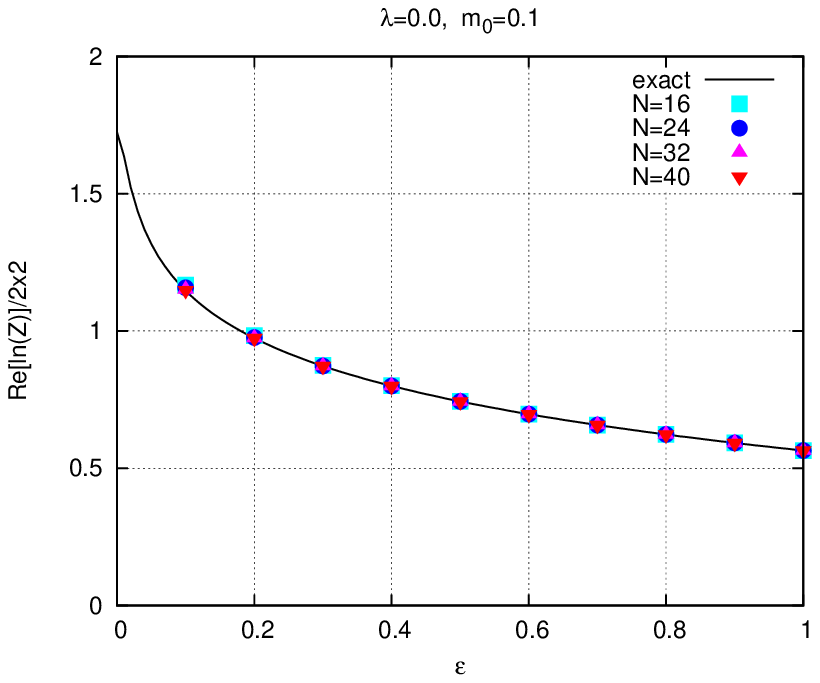}}
&
\hspace{-15mm}
\scalebox{0.9}{\includegraphics{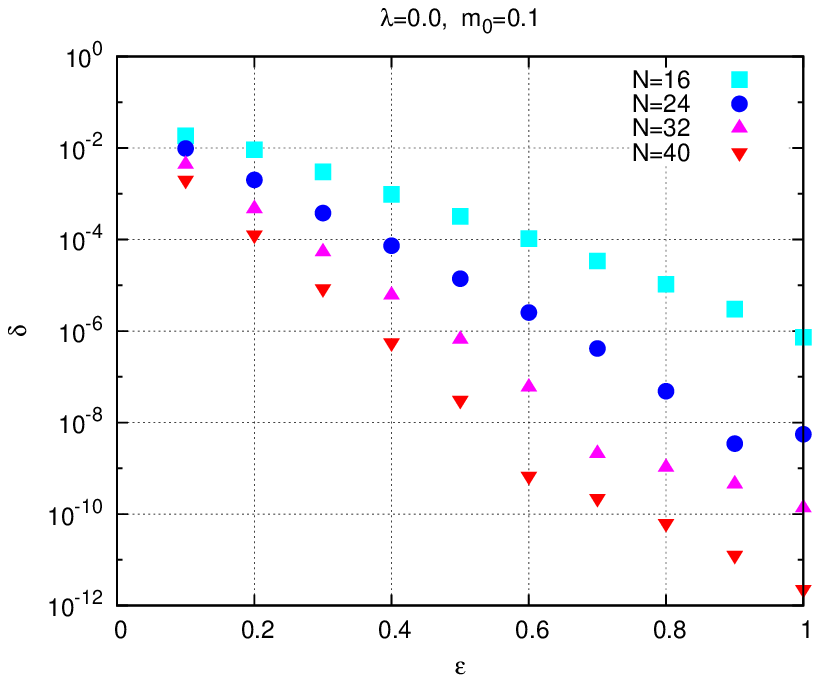}}
\end{tabular}
\end{center}
\vspace{10mm}
\caption{
Left panel: ${\rm Re}[\ln {\cal Z}]/V$ on $2\times2$ lattice as a function of $\varepsilon$ with various
truncation orders $N=16-40$ at $m_0=0.1$ and $\lambda=0$.
The black line is the exact result.
Right panel: The relative deviation for ${\rm Re}[\ln {\cal Z}]$ with the same parameter set as the left one.
}
\label{fig:logZ_2x2}
\end{figure}

In order to check whether the initial tensor in eq.(\ref{eqn:tensor}) is correctly made,
we compute the path integral on $2\times2$ lattice with the periodic boundary condition.
We show the real part of $\ln {\cal Z}/4$ in Fig.~\ref{fig:logZ_2x2} (left).
Here note that we did not use a coarse-graining, thus
this is purely a check of the initial tensor itself.
The exact result and the numerical results with some truncation orders
are consistent with each other at this scale.
Right panel of Fig.~\ref{fig:logZ_2x2} shows
the relative deviation from the exact results
\be
\delta
=
\left|
\frac{{\rm Re}[\ln{\cal Z}_{\rm exact}]-{\rm Re}[\ln {\cal Z}_N]}{{\rm Re}[\ln{\cal Z}_{\rm exact}]}
\right|.
\ee
By increasing $N$, the deviation $\delta$ tends to be smaller.
Note that for smaller $\varepsilon$, the deviation is larger since the
hierarchy of the singular value is not so good as seen before.

\begin{figure}[t]
\begin{center}
\begin{tabular}{cc}
\scalebox{0.9}{\includegraphics{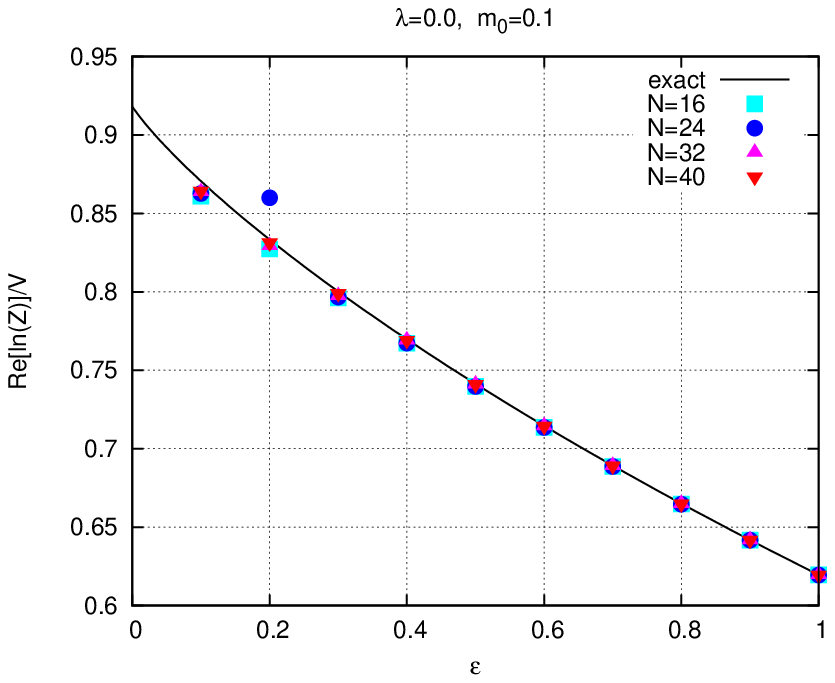}}
&
\hspace{-15mm}
\scalebox{0.9}{\includegraphics{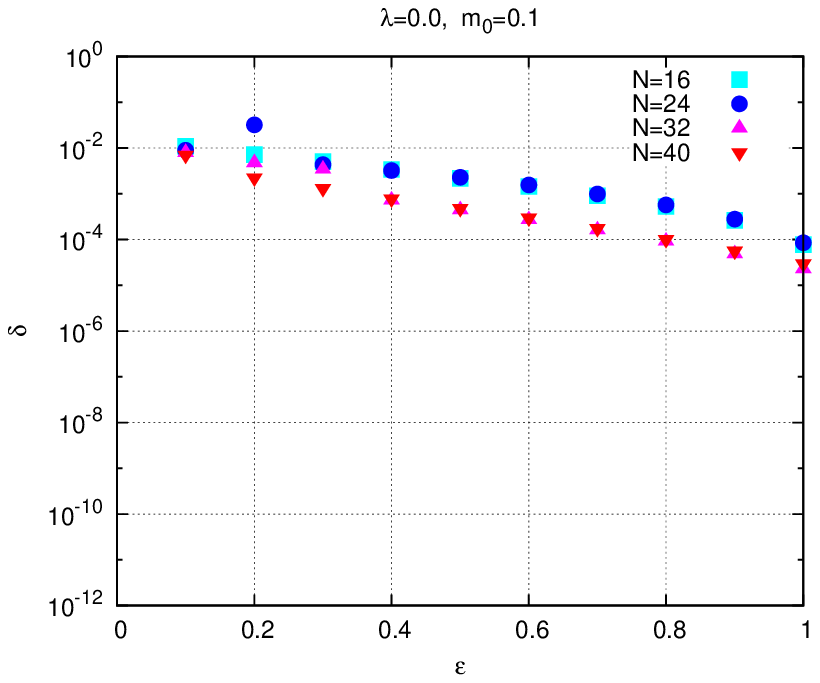}}
\end{tabular}
\end{center}
\vspace{10mm}
\caption{
Left panel: ${\rm Re}[ \ln {\cal Z}]/V$ on $(1024)^2$ lattice as a function of $\varepsilon$. 
Right panel: The associated relative deviation.
The parameter set is the same as that of Fig.~\ref{fig:logZ_2x2}.
}
\label{fig:logZ_V}
\end{figure}

The path integral at larger volumes can be obtained by
using the tensor renormalization group algorithm \cite{Levin:2006jai}.
The results in Fig.~\ref{fig:logZ_V} show that
the coarse-graining step introduces visible errors.
This part should be improved by using some sophisticated algorithms in future .

\section{Summary and outlook}

We have derived the tensor network representation for the scalar field theory with the Minkowskian metric
and numerically checked the validity of the formulation for the free case.
One of the important point in the procedure is to expand the
local factor $H^{(\mu)}$ in terms of the orthonormal basis functions. 
The another key point is to introduce the Feynman prescription $\epsilon$ 
that not only provides
a damping factor in the the path integral but also
dictates the hierarchy of the singular values of $H^{(\mu)}$.
Note that although there is no sign problem in the formulation here,
another problem emerges as an "information incompressibility problem"
which means that the hierarchy of the singular value tends to be worse for smaller $\epsilon$.
Furthermore we have to eliminate the regulator ($\epsilon\to0$) in the end of the calculation.
This is what we have to pay a price in our formulation.

As future outlooks, there are many things to do.
For example, when the $\epsilon$ is small, the initial tensor
has large truncation errors thus one may have to improve the initial tensor itself.
Secondly,  when making the initial tensor for the interacting case,
one needs an efficient numerical method to evaluate the oscillating integral,
say the steepest descent method and so on.
Thirdly, instead of the Feynman prescription, one may use the tilted time axis as a regulator.
Finally, one may straightforwardly extend to the Euclidean space as well as
the Schwinger-Keldysh formulation in a similar way.
Once such a setup is formulated, the transportation coefficients will
be accessible. 


\vspace{3mm}
This work is supported in part by
JSPS KAKENHI Grant Numbers JP17K05411
and
MEXT as ``Exploratory Challenge on Post-K computer'' (Frontiers of Basic Science: Challenging the Limits).

\end{document}